\documentclass[a4paper,twocolumn,showpacs,amsmath,pra,aps,amssymb,nofootinbib,superscriptaddress]{revtex4-1}

\usepackage[T1]{fontenc}
\usepackage[latin9]{inputenc} 
\usepackage{times}
\usepackage{amssymb,amsmath}
\usepackage[pdftex]{graphicx}

\usepackage{subfig}

\graphicspath{{figures/}}

\newcommand{\xv}{\mathbf{x}}
\newcommand{\hc}{^{\dagger}}
\newcommand{\dd}[1]{d#1\,}
\newcommand{\tss}[1]{\textsuperscript{#1}}

\newcommand{\br}{\mathbf{r}}
\newcommand{\Cdd}{C_{\mathrm{dd}}}

\begin{document}

\title{Finite Temperature Dipolar ultra-cold Bose gas with Exchange Interactions}

\author{S. C. Cormack}
\affiliation{Jack Dodd Centre for Quantum Technology, Department of Physics, University of Otago, Dunedin, New Zealand}
\author{D. A. W. Hutchinson}
\affiliation{Jack Dodd Centre for Quantum Technology, Department of Physics, University of Otago, Dunedin, New Zealand}
\affiliation{Centre for Quantum Technologies, National University of Singapore, 3 Science Drive 2, Singapore 117543}

\date{\today}

\begin{abstract}
We develop finite temperature theory for a trapped dipolar Bose gas including thermal exchange interactions. Previous treatments neglected these, difficult to compute, terms. We present a methodology for numerically evaluating the thermal exchange contributions, making use of cylindrical symmetry. We then investigate properties of the dipolar gas, including calculating the excitation spectrum over the full range of trap anisotropy. We evaluate the contributions due to thermal exchange noting that, under some regimes, these effects can be at least as significant as the direct interaction. We therefore provide guidance as to when these cumbersome terms can be neglected and when care should be exercised regarding their omission. 
\end{abstract}

\pacs{03.75.Hh, 64.60.My}
\maketitle

\section{Introduction}
The creation of a Bose-Einstein condensate (BEC) of \tss{52}Cr atoms in 2005 \cite{Griesmaier2005a} has sparked many experimental and theoretical investigations of the effect of long range dipole interactions on BECs. More recently, both \tss{164}Dy \cite{Lu2011a} and \tss{168}Er \cite{Aikawa2012a} have been Bose-condensed. These three atomic species have large magnetic dipole moments compared to alkali atoms. There has also been progress towards a condensed dipolar Bose gas of molecules \cite{Aikawa2010a} which gives the potential for much larger dipole moments. A recent experiment \cite{Bismut2010a} has explored the effect of dipole interactions on collective excitations. Theoretical investigation of the effect of dipolar interactions has often been focused on zero temperature \cite{Santos2000a,Santos2003a,ODell2004a,Ronen2006a}. These works have studied excitations of the dipolar condensate as well as its ground state properties. Previous approaches to studying finite temperature effects include path integral Monte Carlo simulations \cite{Nho2005a,Filinov2010a} and Hartree-Fock-Bogoliubov theory \cite{Ronen2007b}.

We usually consider atoms in a BEC to interact via a van der Waals potential which drops off like $1/r^6$ at large distances. It can be shown that for such a potential, the low energy scattering is dominated by the $s$-wave scattering \cite{Lahaye2009a}. At temperatures low enough for a BEC to form, only low energy scattering occurs so we can safely replace the actual interaction potential with a pseudopotential which reproduces the $s$-wave scattering. This pseudopotential is usually a delta function with the same $s$-wave scattering length. We cannot always assume that interactions between particles in a BEC are short-range only however. Chromium, dysprosium and erbium have significant magnetic dipole moments ($6\mu_B$ for \tss{52}Cr \cite{Griesmaier2005a}, $10\mu_B$ for \tss{164}Dy \cite{Lu2011a} and $7\mu_B$ for \tss{168}Er \cite{Aikawa2012a}, where $\mu_B$ is the Bohr magneton). The dipolar interaction potential drops off like $1/r^3$ for large distances. In this case, all of the higher order partial waves contribute equally to the scattering at low energy \cite{Lahaye2009a}. We therefore cannot replace the true potential with a short-range pseudopotential. When atoms with significant magnetic dipole moments are Bose-condensed we must take into account the long-range nature of the dipolar interaction between atoms and this gives rise to new and interesting behaviour.

Yi and You \cite{Yi2000a}, and G\'oral \textit{et al.}~\cite{Goral2000a} were the first to calculate the zero temperature density profiles for dipolar BECs using a dipolar Gross-Pitaevskii equation (GPE). Yi and You looked at the case of a pancake trap  with $\lambda=\sqrt{8}$. They found that increasing the strength of dipolar interaction caused the condensate to contract in the radial direction and expand in the axial direction while the peak density increased. Meanwhile, G\'oral \textit{et al.}~compared the density profile of the dipolar BEC to an equivalent BEC with only contact interactions. They found that for a pancake trap, the dipolar BEC was larger in the radial direction. For a cigar trap, the condensate was smaller in both the radial and axial direction. These studies showed that the anisotropic nature of the dipole interaction had significant effects on the shape of a condensate. The dipolar GPE was also used in the work of Santos \textit{et al.}~\cite{Santos2000a} to study the stability of a dipolar condensate with no contact interactions.

The natural next direction is to look at the excitation spectrum of the system. Again, this problem was first approached by Yi and You \cite{Yi2002a}, and G\'oral and Santos \cite{Goral2002b}. Both groups used a time-dependent, Gaussian variational ansatz in the GPE to calculate the energies of the three lowest excitations. Both groups also compare these results to numerical solutions of the time-dependent GPE. They do this by starting with the ground state wavefunction and applying a time-varying potential to excite oscillations. Fourier analysis can then be applied to the changing condensate width to determine the oscillation frequency components. Both groups find good agreement between variational and numerical results when the dipolar interaction is relatively weak, however, G\'oral and Santos find that this agreement breaks down when the dipolar interactions start to drive the condensate towards collapse. 

The standard approach to determining the excitation energies of a BEC is to use Bogoliubov theory \cite{Pitaevskii2003a}. This approach was used by Ronen \textit{et al.}~\cite{Ronen2006a} to calculate the excitations of the dipolar BEC. The Bogoliubov method was not used by previous studies \cite{Yi2002a,Goral2002b} due to the numerical difficulty in solving the equations. Ronen \textit{et al.}~overcame this difficulty by utilising both the cylindrical symmetry of the problem and the convenient form of the dipolar interaction term in momentum space. The authors calculated the excitation energies for varying dipole strengths in a number of trap geometries. They also calculated the quantum depletion of the condensate.

In another article, Ronen \textit{et al.}~\cite{Ronen2007a} examined the stability of the dipolar gas. They found that, contrary to a previous study \cite{Santos2000a}, the condensate is unstable for any trap geometry if enough particles are added. They also found that for highly pancake traps with particular aspect ratios (e.g.\ $\lambda\approx 7$), the condensate develops a biconcave, or red blood cell-like, shape as it approaches instability. Ronen and Bohn \cite{Ronen2007b} extended their method to condensates at non-zero temperature using Hartree-Fock-Bogoliubov theory with the Popov approximation (HFBP) \cite{Popov1987a,Griffin1996a}. The excitation energies and condensate fraction are calculated as a function of temperature. They also found that biconcave condensates still existed at finite temperature. 

In their finite temperature treatment, Ronen and Bohn ignore the effect of dipolar exchange from the thermal cloud on the condensate. The associated term in the HFBP equations is difficult to evaluate and was believed to be small. Here we include this term in order to determine whether it has any significant effects. In section \ref{sec:Formalism} we will review the formalism of dipolar HFBP and the algorithm developed in earlier studies \cite{Ronen2006a,Ronen2007b} to solve the associated equations for a cylindrically symmetric system. We will show how to calculate the thermal dipolar exchange term within this method. In sections \ref{sec:Density} and \ref{sec:Excitations} we calculate the density profiles and excitation energies for the system without including the thermal dipolar exchange term. In section \ref{sec:Exchange} we then show the effect that the exchange term has on these properties.

\section{Formalism\label{sec:Formalism}}
We describe the Bose condensed system using HFBP. This is the approach used by the authors of \cite{Ronen2007b}. In that work, dipolar exchange interactions from the thermal cloud acting on the condensate were ignored due to the difficulty in evaluating such terms. Here we calculate these terms in order to estimate how significant an effect they have on the properties of a condensate. The system is assumed to be in thermal equilibrium in the grand canonical ensemble. We work below the critical temperature and fix $N_0$ atoms in the condensate. The chemical potential, $\mu$, and the total number of atoms, $N$, are then determined by the temperature $T$.

We examine a cylindrically symmetric system in a harmonic trapping potential, $V_{\mathrm{tr}}(\xv) =\frac{M}{2}(\omega_\rho^2\rho^2 + \omega_z^2z^2)$, where $M$ is the atomic mass. 
We consider a gas of bosons that interact by both a long-range dipole-dipole interaction and a contact interaction characterized by $V(\br)=g\delta(\br)+V_\mathrm{dd}(\br)$, where $g=4\pi a\hbar^2/M$, with $a$ the $s$-wave scattering length. We take the dipoles to be polarized along $z$ giving a dipole interaction of
\begin{align}
V_\mathrm{dd}(\br)=\frac{\Cdd}{4\pi}\frac{1-3\cos^2\theta}{|\br|^3},\label{e:udd}
\end{align}
where $\Cdd=\mu_0\mu^2_m$ for magnetic dipoles of strength $\mu_m$ and $d^2/\epsilon_0$ for electric dipoles of strength $d$, and $\theta$ is the angle between $\br$ and the $z$ axis.

The grand canonical Hamiltonian for this system is
\begin{multline}
\label{eq:Kfull}
\hat{K} = \int\dd{\xv}\hat{\Psi}\hc(\xv)\left(\hat{H}_{\mathrm{sp}}-\mu\right)\hat{\Psi}(\xv) \\+ \frac{1}{2}\int\dd{\xv}\dd{\xv'}\hat{\Psi}\hc(\xv)\hat{\Psi}\hc(\xv')V(\xv'-\xv)\hat{\Psi}(\xv')\hat{\Psi}(\xv)
\end{multline}
where $\hat{\Psi}(\xv)$ is the usual Bose field operator and $\hat{H}_{\mathrm{sp}}=(-\hbar^2/2M)\nabla^2 + V_\mathrm{tr}(\xv)$ is the single particle Hamiltonian.

The HFBP method consists of expanding the field operator in a series of basis states and replacing the ground state operator with a c-number representing the condensate. We write this as $\hat{\Psi}(\xv) = \sqrt{N_0}\phi_0(\xv) + \tilde{\psi}(\xv)$, where $\phi_0(\xv)$ is the condensate wavefunction and $\tilde{\psi}(\xv)$ is the fluctuation operator representing the other modes. When this is substituted into Eq.~\eqref{eq:Kfull} we can group the resulting terms based on the number of factors of $\tilde{\psi}(\xv)$ they contain. This gives terms up to quartic in $\tilde{\psi}(\xv)$, however, in the HFBP approach, the cubic and quartic terms are approximated using a mean-field factorisation giving terms which are quadratic or lower. We may therefore write the Hamiltonian as $\hat{K} = \hat{K}_0 + \hat{K}_1 + \hat{K}_2$. By making the Hamiltonian stationary with respect to arbitrary first-order variation in $\tilde{\psi}(\xv)$, we obtain the generalised GPE for $\phi_0(\xv)$
\begin{multline}
\label{eq:GPE}
\left[\hat{H}_\mathrm{sp}+gn_c(\xv) + 2g\tilde{n}(\xv) + \Phi_D(\xv)\right]\phi_0(\xv) \\+ \Phi_E[\phi_0(\xv)] =\mu\phi_0(\xv)
\end{multline}
where $g=4\pi\hbar^2a/M$, $n_c(\xv)\equiv N_0|\phi_0(\xv)|^2$ is the condensate density, and $\tilde{n}(\xv)\equiv\langle\tilde{\psi}\hc(\xv)\tilde{\psi}(\xv)\rangle$ is the non-condensate density. The terms involving $g$ come from the contact interaction, while $\Phi_D$ and $\Phi_E$ arise from the dipole-dipole interaction. The term $\Phi_D(\xv)$ represents the direct dipole interaction and has the form $\Phi_D(\xv)=\int\dd{\xv'}V_\mathrm{dd}(\xv'-\xv)[n_c(\xv')+\tilde{n}(\xv')]$. The other dipole term, $\Phi_E[\phi_0(\xv)]$ represents the dipole exchange interaction between the condensate and non-condensate. It may be calculated from
\begin{equation}
\label{eq:PhiE}
\Phi_E[\phi_0(\xv)] = \int\dd{\xv'}\tilde{n}(\xv',\xv)V_\mathrm{dd}(\xv'-\xv)\phi_0(\xv')
\end{equation}
where $\tilde{n}(\xv',\xv)\equiv\langle\tilde{\psi}\hc(\xv')\tilde{\psi}(\xv)\rangle$ is the non-condensate one body density matrix.

In order to determine the excitations and non-condensate properties we turn to the second order  contribution to the Hamiltonian, $\hat{K}_2$. This may be diagonalised by applying a Bogoliubov transformation to the fluctuation operator, transforming to a new set of bosonic quasiparticle operators,
\begin{equation}
\label{eq:BogTrans}
\tilde{\psi}(\xv) = \sum_j \left[ u_j(\xv)\alpha_j-v_j^*(\xv)\alpha_j\hc \right]
\end{equation}
and requiring $u_j(\xv)$ and $v_j(\xv)$ to satisfy the Bogoliubov-de Gennes (BdG) equations
\begin{equation}
\begin{gathered}
\label{eq:HFBP}
\hat{\mathcal{L}}u_j(\xv) - \hat{M}v_j(\xv) = E_ju_j(\xv)\\
\hat{\mathcal{L}}^*v_j(\xv) - \hat{M}^*u_j(\xv) = -E_jv_j(\xv)
\end{gathered}
\end{equation}
where $\hat{\mathcal{L}}=\hat{h}_{0}-\mu+\hat{M}$. $\hat{h}_{0}$ is the operator on the left hand side of Eq.~\eqref{eq:GPE}, i.e.\ $\hat{h}_{0}\phi_0(\xv)=\mu\phi_0(\xv)$, while $\hat{M}$ is defined such that for an arbitrary function $\psi(\xv)$,
\begin{multline}
\label{eq:M}
\hat{M}\psi(\xv) = gn_c(\xv)\psi(\xv) +\\
 N_0\phi_0(\xv)\int\dd{\xv'}\phi_0(\xv')V_\mathrm{dd}(\xv'-\xv)\psi(\xv')
\end{multline}
Terms involving $\hat{M}$ represent exchange interaction with the condensate. The thermal one body density matrix can then be evaluated from
\begin{align}
    \tilde{n}(\xv',\xv) = \sum_j &\left\{\left[u_j^*(\xv')u_j(\xv)+v_j(\xv')v_j^*(\xv)\right]\overline{n}_\mathrm{BE}(E_j) \right.\notag\\&\left.+ v_j(\xv')v_j^*(\xv)\right\}
\label{eq:nT1bdm}
\end{align}
where $\overline{n}_\mathrm{BE}$ is the Bose distribution for the quasiparticles. The non-condensate density is simply $\tilde{n}(\xv)=\tilde{n}(\xv,\xv)$.

Following the method of \cite{Hutchinson1997a}, to solve the BdG equations we first decouple equations \eqref{eq:HFBP} by introducing the new amplitudes $\psi_j^{\pm}(\xv)=u_j(\xv)\pm v_j(\xv)$. In terms of these amplitudes, the BdG equations are
\begin{subequations}

\begin{align}
\label{eq:HFBPpma}
(\hat{h}_0-\mu+2\hat{M})(\hat{h}_0-\mu)\psi_j^{+}(\xv) &= E_j^2\psi_j^{+}(\xv)\\
\label{eq:HFBPpmb}
(\hat{h}_0-\mu)(\hat{h}_0-\mu+2\hat{M})\psi_j^{-}(\xv) &= E_j^2\psi_j^{-}(\xv)
\end{align}
\end{subequations}
Either of these equations can be used to determine the excitation energies, $E_j$, and original amplitudes $u_j(\xv)$ and $v_j(\xv)$. We choose to solve the first. We do this by expanding $\psi_j^{+}(\xv)$ in the basis of eigenstates of the GPE \eqref{eq:GPE}, excluding the ground state, $\phi_0(\xv)$, i.e.\ we write $\psi_j^{+}(\xv)=\sum_\alpha c^j_\alpha \phi_\alpha(\xv)$, where $(\hat{h}_0-\mu)\phi_\alpha(\xv)=\epsilon_\alpha\phi_\alpha(\xv)$. The eigenvalues, $\epsilon_\alpha$ are just the eigenvalues of the GPE relative to the chemical potential. This method has the advantage of ensuring that the excitations are orthogonal to the condensate. In this basis, Eq.~\eqref{eq:HFBPpma} is
\begin{equation}
\label{eq:HFBP_GPbasis}
\sum_\alpha\left(\epsilon_\alpha\delta_{\gamma\alpha} + 2M_{\gamma\alpha}\right)\epsilon_\alpha c_\alpha^j = E_j^2 c_\gamma^j
\end{equation}
where $M_{\gamma\alpha}=\int\dd{\xv}\phi_\gamma^*(\xv)\hat{M}\phi_\alpha(\xv)$.

To solve Eq.~\eqref{eq:GPE} we use the algorithm of \cite{Ronen2006a}. With this method, the kinetic and dipolar interaction terms are calculated using Fourier transforms. For example, the direct dipolar interaction is found from
\begin{equation}
\label{eq:PhiDconv}
\Phi_D(\xv) = \mathcal{F}^{-1}\left[\tilde{V}_\mathrm{dd}(\mathbf{k})(n_c(\mathbf{k})+\tilde{n}(\mathbf{k}))\right]
\end{equation}
by using the convolution theorem. $\mathcal{F}^{-1}$ is the inverse Fourier transform and $\tilde{V}_\mathrm{dd}(\mathbf{k})$, $n_c(\mathbf{k})$ and $\tilde{n}(\mathbf{k})$ are the Fourier transforms of the respective position space quantities. This method allows us to avoid the singular behaviour of $V_\mathrm{dd}(\xv)$ at the origin.  To improve accuracy, we truncated the interaction potential in real space, using the method of \cite{Lu2010a} where necessary.

The cylindrical symmetry of the problem is used by performing the angular integrals analytically and thus converting the 2D Fourier transform in the $x$-$y$ plane into a 1D Hankel transform. The eigenstates of the GPE can be assumed to have the form $f(\xv)=e^{im\phi}f(\rho,z)$ where $m$ is an integer. Their Fourier transform, and any quantities with the same angular dependence, can be reduced to
\begin{equation}
\label{eq:FHf}
\tilde{f}(k_\rho,k_\phi,k_z) = 2\pi i ^{-m}e^{imk_\phi}\int\dd{\rho}\rho f(\rho,z)J_m(k_\rho \rho)
\end{equation}
where $J_m(x)$ is the $m$th order Bessel function. We then define the $m$th order 2D Fourier-Hankel transform by
\begin{equation}
\mathcal{F}_m[f(\rho,z)] = 2\pi\int\dd{\rho}\rho f(\rho,z)J_m(k_\rho \rho).
\end{equation}

Previously, the thermal exchange term, $\Phi_E$, in the GP and BdG equations has been ignored as it is computationally intensive to calculate and was not believed to be significant \cite{Ronen2007b}. Here we calculate this term to determine the size of its effect. We can put Eq.~\eqref{eq:PhiE} in a similar form to \eqref{eq:PhiDconv}
\begin{equation}
\label{eq:PhiE_k}
\Phi_E\left[\phi_0(\xv)\right] = \frac{1}{(2\pi)^3}\int\dd{\mathbf{k}}e^{i\mathbf{k}\cdot\xv}\tilde V_\mathrm{dd}(\mathbf{k})f(\xv,\mathbf{k})
\end{equation}
where $f(\xv,\mathbf{k})=\int\dd{\xv'}e^{-i\mathbf{k}\cdot\xv'}\tilde n(\xv',\xv)\phi_0(\xv')$. Unfortunately, Eq.~\eqref{eq:PhiE_k} is not a Fourier transform due to the inseparable $\xv$ dependence in $f(\xv,\mathbf{k})$, so the integral must be carried out for each grid position. To calculate excitations with $m>0$ we need to find $\Phi_E[f(\xv)]$ for $f(\xv)=e^{im\phi}f(\rho,z)$. The Bogoliubov excitations correspond to distinct $m$ values so we may write $u_j(\xv)=e^{im\phi}u_{jm}(\rho,z)$ and likewise for $v_j(\xv)$. 
Carrying out all of the angular integrals in Eq.~\eqref{eq:PhiE_k} analytically, we get an expression for $\Phi_E$ which can be evaluated on a $(\rho,z)$ grid
\begin{align}
&\Phi_E[e^{im\phi}f(\rho,z)] \notag\\ &= \frac{e^{im\phi}}{2\pi}\int\dd{k_\rho}k_\rho\int\dd{k_z}e^{ik_{z}z}\tilde{V}_\mathrm{dd}(k_\rho,k_z)\tilde{n}^E_m(\rho,z,k_\rho,k_z),
\label{eq:PhiE_rz}
\end{align}
where
\begin{align}
    \hspace{-1mm}   \tilde{n}_m^E(\rho,z,k_\rho,k_z) &=\hspace{-2mm} \sum_{m'=0}^\infty\hspace{-1.5mm} \mathcal{F}_{|m'-m|}[\tilde n_{m'}(\rho',z',\rho,z)]J_{|m'-m|}(k_{\rho}\rho)\notag\\
 &+ \sum_{m'=1}^\infty \hspace{-1.5mm}\mathcal{F}_{m'+m}[\tilde n_{m'}^*(\rho',z',\rho,z)]J_{m'+m}(k_{\rho}\rho).
\label{eq:Gsum}
\end{align}
and
\begin{align}
 \tilde n_{m}&(\rho',z',\rho,z) = \notag\\
 \sum_j&\left\{\left[u_{jm}^*(\rho',z')u_{jm}(\rho,z) + v^*_{jm}(\rho',z')v_{jm}(\rho,z)\right]
 \bar{n}_\mathrm{BE}(E_j)\right.\notag\\
 &\left.+v^*_{jm}(\rho',z')v_{jm}(\rho,z)\right\}f(\rho',z').
\end{align}
A number of schemes for choosing dimensionless parameters have been used in the literature. The energy and length units are usually chosen based on the harmonic trap frequency. Except where stated, $a_\mathrm{ho} \equiv \sqrt{\hbar/M\omega_\rho}$. We also define \cite{Hutchinson1998a}
\begin{equation}
\label{eq:ariMean}
\omega_a^2 = \frac{1}{3}(2\omega_\rho^2+\omega_z^2).
\end{equation}
To characterise the strength of the dipolar and contact interactions we define the dimensionless parameters, 
\begin{equation}
D=\frac{NM\Cdd}{4\pi\hbar^2a_\mathrm{ho}}
\end{equation}
and
\begin{equation}
    \tilde{g} = \frac{NMg}{\hbar^2a_\mathrm{ho}}=\frac{4\pi Na}{a_\mathrm{ho}}
\end{equation}
where $N$ is the number of particles and $a$ is the scattering length.

\section{Density Profiles\label{sec:Density}}

\begin{figure}
\centering
\includegraphics[scale=1]{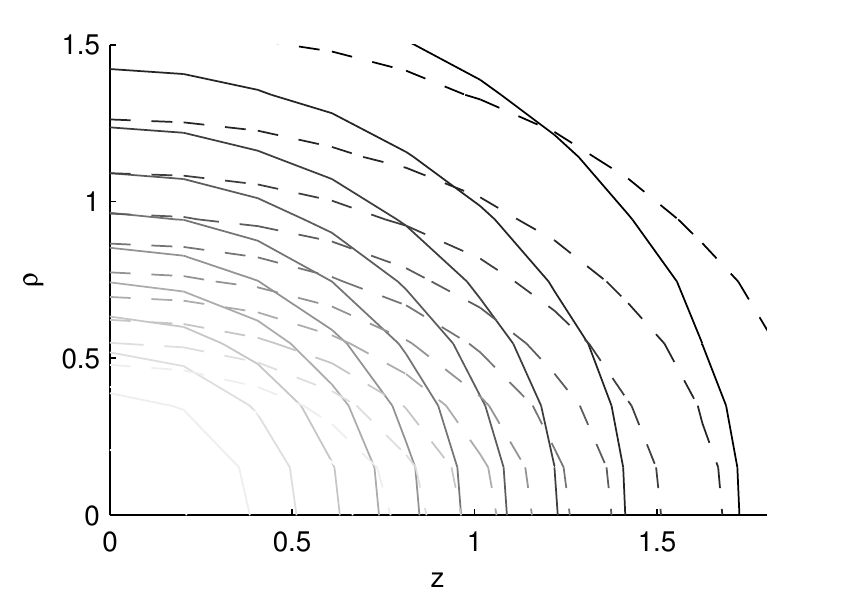}
\caption{Contour plots of condensate density for contact interactions only (solid) and contact and dipole interactions (dashed), with $\lambda=1$, $T=\hbar\omega_\rho/k_B$, $\tilde{g} = 1$, $N=10^4$ and $D=3$ for the dipolar case. The density decreases monotonically from the origin. $\rho,z$ are in units of $a_\mathrm{ho}$.}
\label{fig:ncNoExCompare}
\end{figure}

As a basis for comparison, we first examine the solutions of the HFBP equations without including thermal dipolar exchange. The first quantity we will look at is the condensate density. The presence of dipolar interactions alters the shape of the condensate as was seen in the earliest studies of dipolar condensates \cite{Yi2000a,Goral2000a}. The same behaviour occurs at finite temperature as we can see in Fig.~\ref{fig:ncNoExCompare}. When there are only contact interactions present, a condensate in a spherical trap is also spherical. When there are dipolar interactions present, however, the condensate elongates in the axial direction.

This behaviour can be explained by considering the anisotropy of the dipolar potential. Dipoles experience an attractive force when they are aligned head-to-tail rather than side-by-side. A density profile which is elongated along the axial direction, so that more dipoles are aligned head-to-tail, will have a lower energy than one in which the dipoles experience more repulsive forces. This is not a particularly intuitive result as one might expect that the repulsive interactions in the $x$-$y$ plane would cause the condensate to expand radially. Since the dipoles are attractive along the axial direction however, they may have a lower energy overall by aligning head-to-tail than they would by simply moving apart. The same behaviour was found at zero temperature by Yi and You \cite{Yi2000a} for a slightly pancake trap.

\begin{figure}
\centering
\subfloat[Cigar trap, $\lambda=1/7$]{\label{fig:ncCigar}\includegraphics[scale=1]{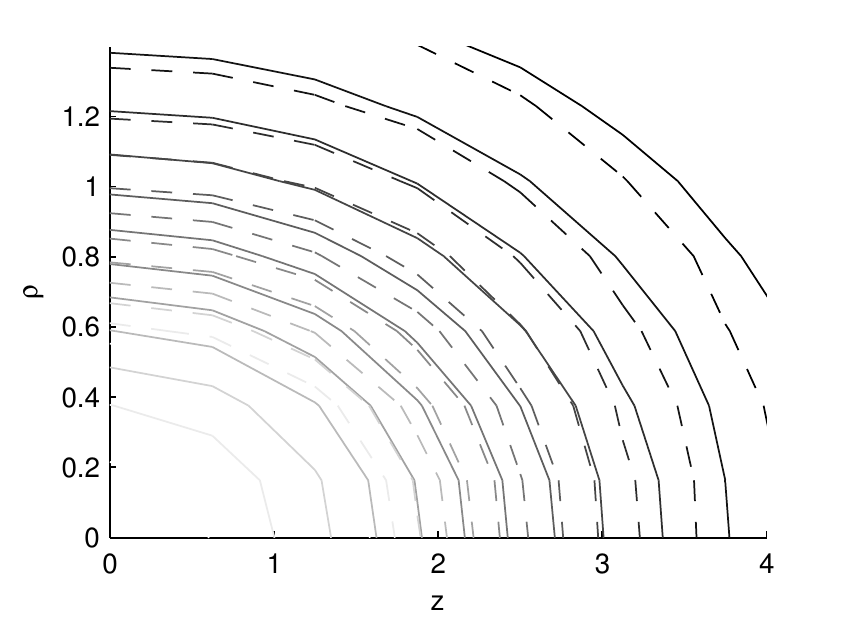}}

\subfloat[Pancake trap, $\lambda=7$]{\label{fig:ncPancake}\includegraphics[scale=1]{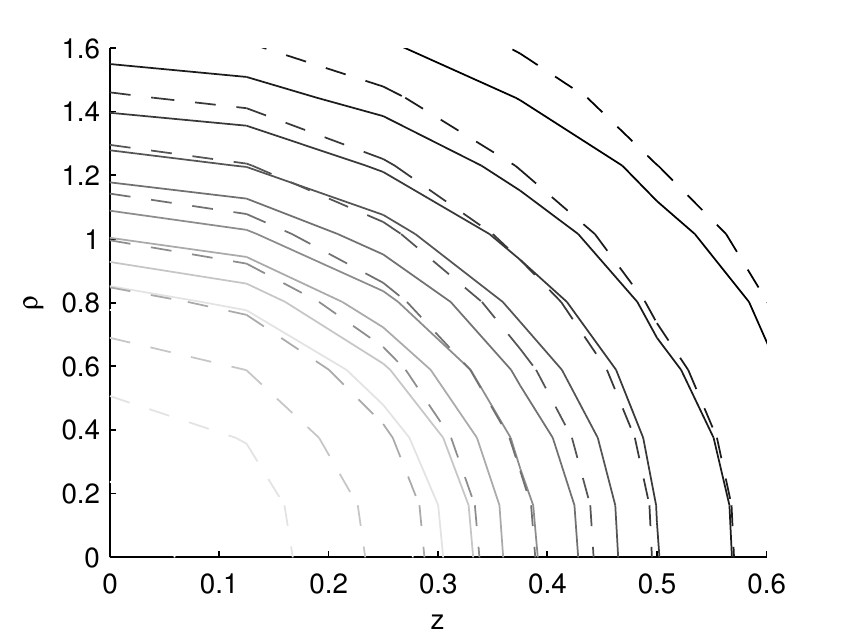}}
\caption{Condensate density with contact interactions only (solid) and with contact and dipolar interactions (dashed). (a) cigar-shaped trap with $\lambda=1/7$ and $T=0.5\hbar\omega_\rho/k_B$, (b) pancake trap with $\lambda=7$ and $T=\hbar\omega_\rho/k_B$. All cases use $N=10^4$ and $\tilde{g}=1$. For the dipolar cases we used $D=3$. The contour lines decrease in density from the centre outwards. Within each subfigure the solid and dashed contour lines are for the same density values. $\rho,z$ are in units of $a_\mathrm{ho}$.}
\label{fig:ncAspect}
\end{figure}

The effect of the dipolar interactions on the shape of the condensate is highly dependent on the aspect ratio of the trap \cite{Goral2000a}. In Fig.~\ref{fig:ncAspect} we have plotted the condensate density contours for a cigar trap with $\lambda=1/7$ and a pancake trap with $\lambda=7$. In the cigar-shaped trap, the condensate will be elongated in the axial direction. This means that the attractive part of the dipolar interaction will be dominant. In Fig.~\ref{fig:ncCigar} we can see that the dipolar condensate drops off more steeply from the centre than the contact-only condensate. It also has a higher peak density. The attractive part of the dipolar interaction is `pulling' the condensate in on itself. In this case the dipolar interactions have little effect on the aspect ratio of the condensate.

In the pancake-shaped trap in Fig.~\ref{fig:ncPancake}(b) we can see that the dipolar interaction affects both the drop-off in density and its aspect ratio. The condensate expands radially with the dipolar interactions and its density decreases more slowly from the centre than the contact-only case. This time the dipolar condensate has a lower peak density than the contact-only condensate. In the pancake trap, the repulsive part of the dipolar interaction is dominant and this explains these effects. Unlike in the spherically trapped situation, the condensate does expand radially due to the repulsion of side-by-side dipoles. In this case, the dipoles face a large energy cost for lining up head-to-tail due to the strong axial trapping. The lowest energy state is therefore when the dipoles spread out radially where there is only weak trapping.

We can also look at the non-condensate density. This is given by setting $\xv'=\xv$ in Eq.~\eqref{eq:nT1bdm} giving,
\begin{equation}
\label{eq:nonCondDensity}
\tilde{n}(\xv) = \sum_i \{[|u_i(\xv)|^2+|v_i(\xv)|^2]\overline{n}_\mathrm{BE}(E_i) + |v_i(\xv)|^2\}.
\end{equation}
At zero temperature, $\overline{n}_\mathrm{BE}(E)=0$ so the only contribution to the non-condensate density is the \emph{quantum depletion}, $\tilde{n}(\xv) = \sum_i|v_i(\xv)|^2$. This term generally requires the calculation of very many modes to converge, however, it is usually very small \cite{Ronen2006a}, and at finite temperature it is overwhelmed by the thermal population of the excited modes. The thermal part of the non-condensate converges more quickly as the Bose distribution factor decays exponentially with the mode energy once $\beta E_i\gg1$. The fact that the quantum depletion may not be well converged is therefore not very important once there is an appreciable amount of thermal atoms.

\begin{figure}
\centering
\subfloat[Contact interactions only]{\label{fig:ntildeNoExCon}\includegraphics[scale=1]{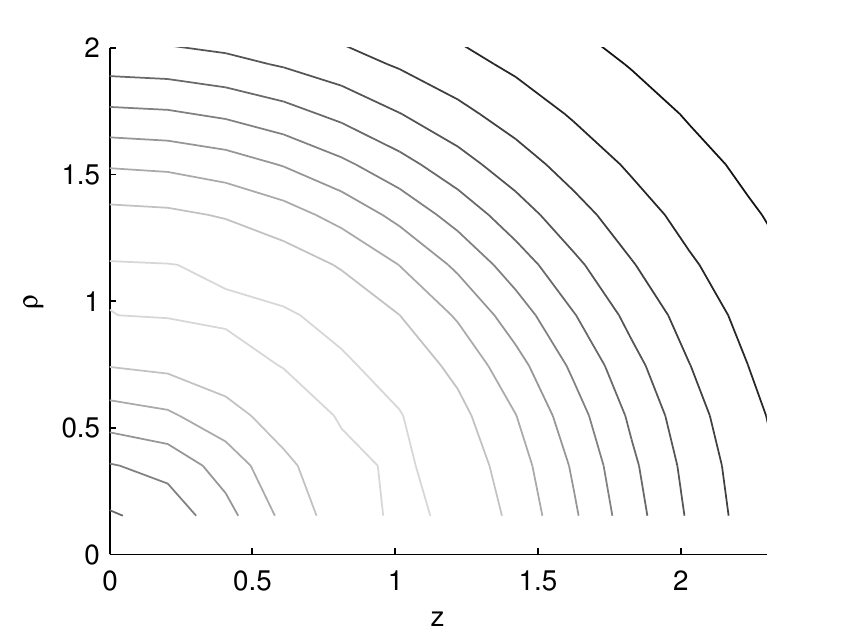}}

\subfloat[Contact and dipole interactions]{\label{fig:ntildeNoExDip}\includegraphics[scale=1]{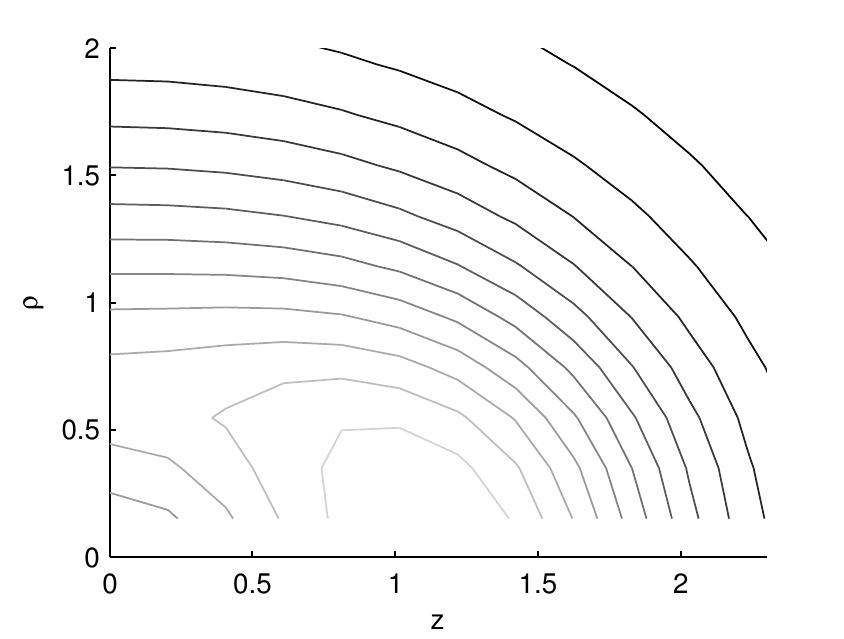}}
\caption{Non-condensate density (a) without and (b) with dipolar interactions. Light is the highest density and dark is the lowest density. Parameters used are the same as for Fig.~\ref{fig:ncNoExCompare}. Some of the contours are jagged due to the limited spatial resolution. $\rho,z$ are in units of $a_\mathrm{ho}$.}
\label{fig:ntildeNoEx}
\end{figure}

Two examples of non-condensate densities are shown in Fig.~\ref{fig:ntildeNoEx}. The temperature used for these calculations is $T = \hbar\omega_\rho /k_B$. This temperature is high enough that the contribution to the non-condensate density from thermal particles is much greater than that from the quantum depletion. It is also low enough that the energy cutoff of $E_{\mathrm{cut}}=6\hbar\omega_\rho$ does not significantly affect the result. Fig.~\ref{fig:ntildeNoExCon} shows the non-condensate density when there are only contact interactions present. The density is spherically symmetric, increasing from the centre of the trap to a maximum at $\rho^2+z^2\approx 1$ and decreasing thereafter. This ring shaped thermal cloud also occurs when no interactions at all are included. This is simply due to the fact that the lowest modes above the condensate mode have peaks away from the centre of the trap. This ring shaped cloud is enhanced by the presence of contact interactions as the thermal cloud experiences a potential of $2gn_c(\xv)$ from the condensate which pushes it out. Fig.~\ref{fig:ntildeNoExDip} shows the non-condensate density with dipolar interactions included. Compared with the non-dipolar case, the density is reduced in the $z=0$ plane of the trap and there are two separate peaks at $z/a_\mathrm{ho}\approx\pm 1$. The two peaks arise due to the anisotropic dipolar interaction splitting up the ring from the contact interaction case.

In this section we have shown the effects of dipolar interactions on the finite temperature BEC. The interactions can have a significant impact on the aspect ratio of the BEC, breaking the spherical symmetry when the trap is spherically symmetric. The effect of the interactions is strongly dependent on the aspect ratio of trap as this determines whether the attractive or repulsive part of the dipolar potential is dominant. In section \ref{sec:Exchange} we will include the thermal dipolar exchange term and determine what further effect this has on the density of the condensate.

\section{Excitations\label{sec:Excitations}}
We will now look at the excitation spectrum of the dipolar BEC. This is the set of eigenvalues, $E_j$, obtained from the BdG equations \eqref{eq:HFBP}. Previously, the zero temperature Bogoliubov excitations have been studied by Ronen \textit{et al.}~\cite{Ronen2006a}. Ronen and Bohn also looked at the finite temperature excitations without including thermal exchange \cite{Ronen2007b}. They found that the excitation energies are not significantly affected by temperature except near the critical temperature or for highly non-spherical traps (e.g.\ $\lambda=7$). The temperatures we can accurately model are limited by the cutoff in energy we use for the Bogoliubov energies. Over this range of temperatures we also find very little change in excitation energies. We will therefore examine the zero-temperature spectrum as it is significantly easier to calculate and displays all of the same features as the small but finite temperature spectrum. We will also ignore the quantum depletion as its effect is very small and requires many modes to calculate accurately.

\begin{figure*}
\subfloat[Even excitations, $D=3, \tilde{g}=1$]{\includegraphics[scale=0.63]{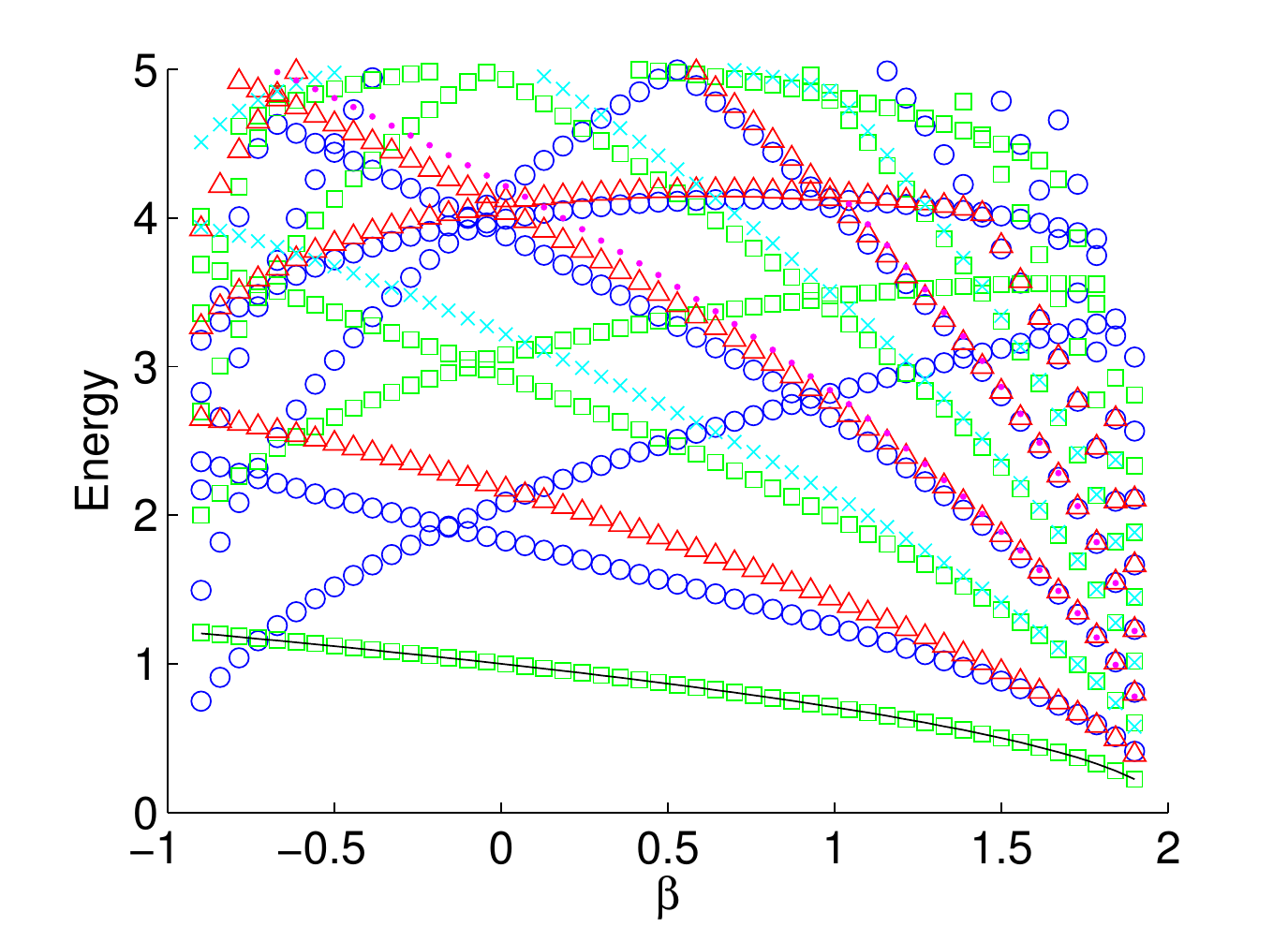}}
\subfloat[Odd excitations, $D=3, \tilde{g}=1$]{\includegraphics[scale=0.63]{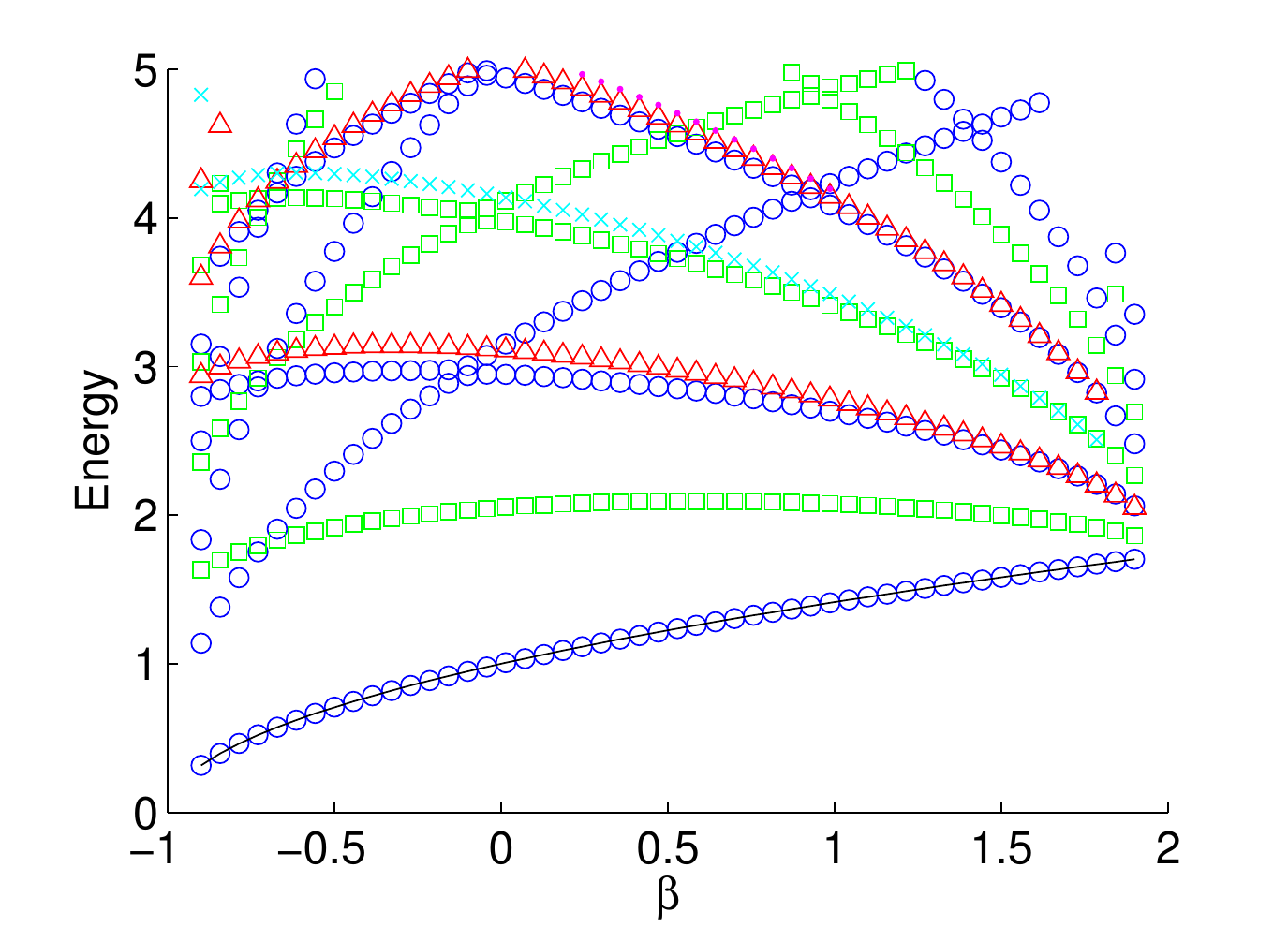}}

\subfloat[Even excitations, $D=0, \tilde{g}=0$]{\includegraphics[scale=0.63]{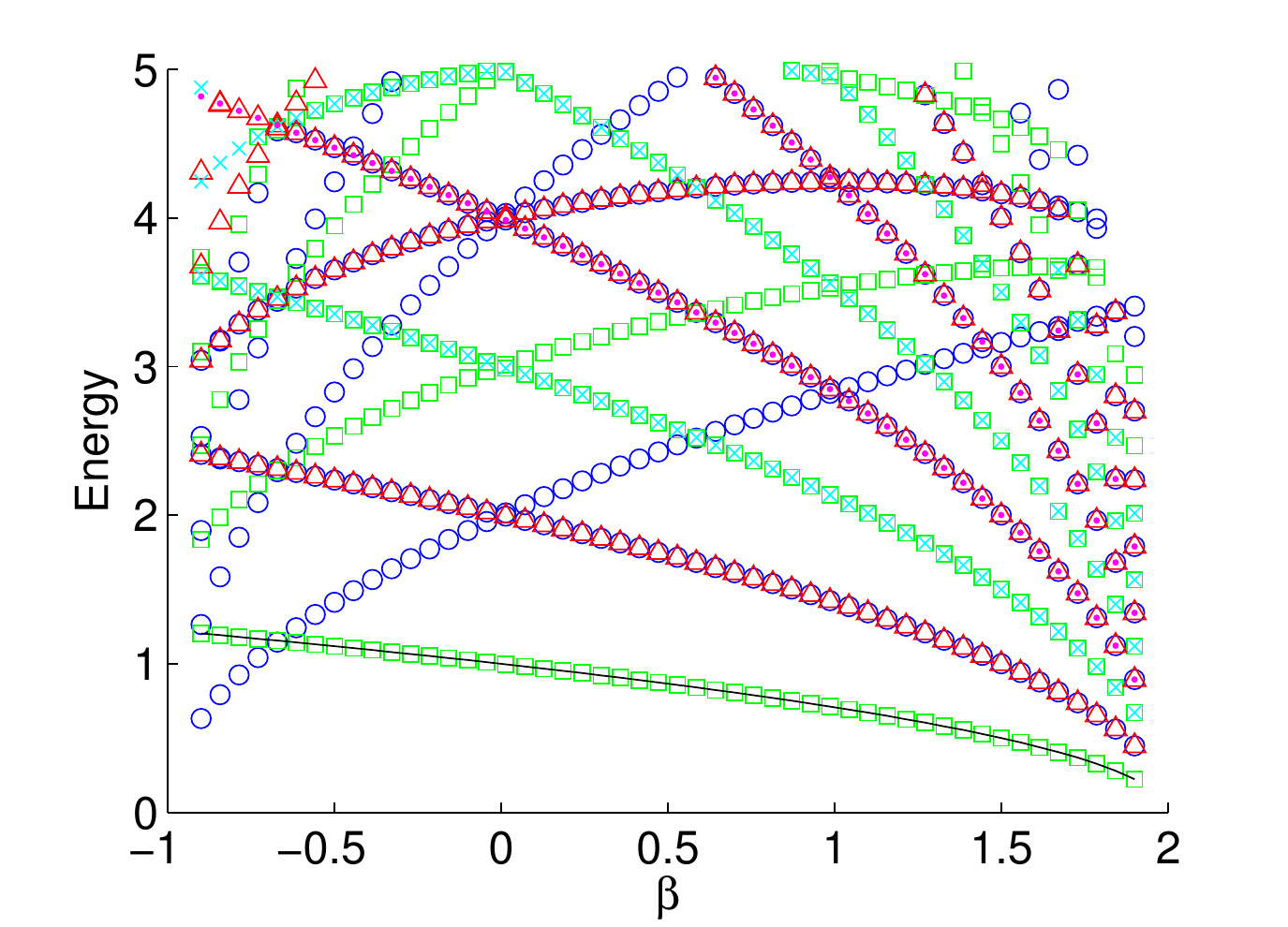}}
\subfloat[Odd excitations, $D=0, \tilde{g}=0$]{\includegraphics[scale=0.63]{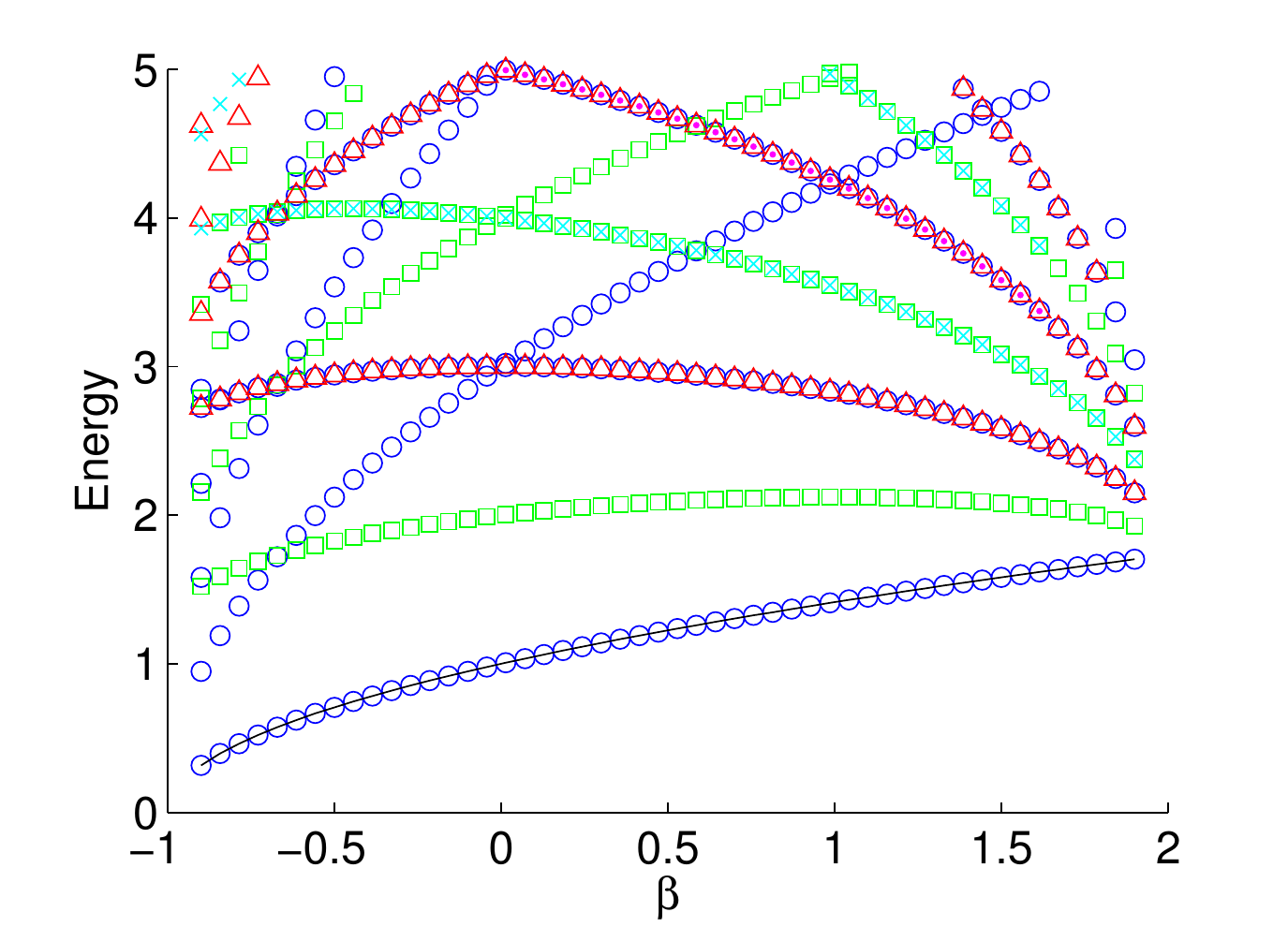}}
\caption{(Color online) Zero temperature excitation energies in units of $\hbar\omega_a$ ($D$ and $\tilde{g}$ use $a_\mathrm{ho}=\sqrt{\hbar/M\omega_a}$). Different $m$ states are represented by blue circles ($m=0$), green squares ($m=1$), red triangles ($m=2$), cyan crosses ($m=3$), and pink dots ($m=4$). The solid black lines show the analytical values for the Kohn modes.}
\label{fig:ExcitsZeroT}
\end{figure*}

First we examine how the excitations are affected by the aspect ratio of the trap. In order to concisely show the full range of trap aspect ratios, we follow the conventions used by Hutchinson and Zaremba \cite{Hutchinson1998a}. The aspect ratio is characterised by the parameter,
\begin{equation}
\beta = \frac{2\omega_z^2-2\omega_\rho^2}{2\omega_\rho^2+\omega_z^2},
\end{equation}
and energy and length scales are defined by the arithmetic mean as defined in Eq.~\eqref{eq:ariMean}. The aspect ratio, $\lambda$, is related to $\beta$ by
\begin{equation}
\lambda = \sqrt{\frac{2\beta+2}{2-\beta}}
\end{equation}
The parameter $\beta$ can range from $-1$, corresponding to a one-dimensional cigar trap, to $2$, corresponding to a two-dimensional pancake trap. The trap is spherical when $\beta=0$. Because the trap potential is symmetric on reflection in the $z=0$ plane, the excitations can be split into those with even or odd amplitudes. The excitation energies are plotted against $\beta$ in Fig.~\ref{fig:ExcitsZeroT}.

The lowest $m=1$ even excitation and lowest $m=0$ odd excitation represent the centre of mass or Kohn modes in the radial and axial directions respectively. The Kohn modes should oscillate at the trap frequencies, $\omega_\rho$ for the radial mode and $\omega_z$ for the axial mode, regardless of the interactions present. Their corresponding energies, in units of the mean trap energy, should therefore be given by $\omega_\rho/\omega_a = \sqrt{1-\beta/2}$ and $\omega_z/\omega_a = \sqrt{1+\beta}$ for the radial and axial modes respectively. These analytical expressions have been plotted along with the calculated excitation energies in Fig.~\ref{fig:ExcitsZeroT}. There is excellent agreement between the analytical and numerical results indicating no problems with the model or numerics.

The excitation energies for a non-interacting BEC are also plotted in Fig.~\ref{fig:ExcitsZeroT} for comparison with the interacting case. The non-interacting excitations are simply calculated by setting both $D$ and $\tilde{g}$ to zero. The contact interaction used in the dipolar calculation is relatively small and the excitations produced using only this interaction are only slightly different to the non-interacting case. The differences between the dipolar calculation and the non-interacting case are therefore primarily due to the dipole interaction.

The dipolar interaction breaks the degeneracy between various modes. This happens in two ways. Excitation energies corresponding to different values of $m$ are split over a large range of the trap aspect ratio. Higher $m$ excitations are pushed up relative to lower $m$ excitations. For example, the lowest even $m=2$ excitation is always degenerate with one of the even $m=0$ excitations in the non-interacting case. When dipolar interactions are included, the $m=2$ excitation has a higher energy for most aspect ratios. As $\beta$ approaches 2, and the trap becomes more pancake shaped, this energy splitting is reduced. This pattern breaks down for highly pancake traps both when the dipolar interaction is very strong \cite{Ronen2007a} and at high temperatures \cite{Ronen2007b}.

The dipolar interactions also break the degeneracy of modes of the same angular momentum at the aspect ratios where they would have crossed each other in the non-interacting case. This leads to avoided crossings between excitations of the same angular momentum. The energy differences are quite small making it difficult to see these avoided crossings in Fig.~\ref{fig:ExcitsZeroT}. 

In this section we have seen the effects of the dipolar interactions on the excitation spectrum of the Bose gas at zero temperature. The interaction breaks almost all of the degeneracies present in the low energy excitations for the ideal gas case, leaving only the degeneracy between $+m$ and $-m$ modes. In the next section we will examine the effect of the thermal exchange term on these excitations at finite temperature.

\section{Thermal Exchange\label{sec:Exchange}}
The dipolar thermal exchange term, \eqref{eq:PhiE_rz}, is the most difficult to evaluate and we therefore attempt to minimise the number of times it is evaluated. We do this by first solving equations \eqref{eq:GPE} and \eqref{eq:HFBPpma} self-consistently without this term. This will be a good approximation to the full solution as long as the effect from dipolar thermal exchange is small. We then solve the full set of equations including dipolar thermal exchange using this solution as an initial guess. Assuming that the full solution is close to the initial guess, convergence should be achieved with relatively few self-consistency iterations.

\begin{table}
\begin{tabular}{ c c c c }
\hline
Cutoff energy/$\hbar\omega_\rho$ & 4 & 6 & 8 \\
$|\mu_E/\mu_{\tilde{D}}|$ & 0.82 & 0.92 & 0.99\\ 
\hline
\end{tabular}
\caption{\label{tab:muRatio}Ratio of contributions to the chemical potential from exchange and direct thermal dipolar interactions. Only modes which have an energy below the cutoff energy are included. The parameters used are $N=10^4$, $\lambda=\sqrt{8}$, $T=\hbar\omega_\rho/k_B$, $D=3.5$ and $\tilde{g}=0$.}
\end{table}

To determine the size of the effect of the thermal exchange term we can examine the quantity $\Phi_E[\phi_0(\xv)]$. This gives the interaction of the non-condensate on the condensate via exchange. We cannot define an effective potential for the exchange as we can with the direct dipolar interaction.

To obtain a convenient measure of the importance of the thermal exchange term, we can look at its contribution to the chemical potential \cite{Ticknor2012a}. The chemical potential can be obtained from the GPE by taking the inner product with $\phi_0(\xv)$  giving
\begin{multline}
\label{eq:chemPot}
\mu  = \int\phi_0^*(\xv)\hat{H}_{\mathrm{sp}}\phi_0(\xv) + [gn_c(\xv)+2g\tilde{n}(\xv) \\ +\Phi_D(\xv)]|\phi_0(\xv)|^2 + \phi_0^*(\xv)\Phi_E[\phi_0(\xv)]\dd{\xv}.
\end{multline}
The contribution of thermal exchange is therefore given by
\begin{equation}
\label{eq:muEx}
\mu_E = \int\phi_0^*(\xv)\Phi_E[\phi_0(\xv)]\dd{\xv}.
\end{equation}
To determine the importance of this term we can compare it with the contribution from the direct dipolar interaction from the non-condensate. This is given by
\begin{equation}
\label{eq:muDirect}
\mu_{\tilde{D}} = \int\dd{\xv}|\phi_0(\xv)|^2\int\dd{\xv'}V_{\mathrm{dd}}(\xv'-\xv)\tilde{n}(\xv').
\end{equation}
We do not compare with the full direct dipolar interaction as this will be dominated by condensate-condensate interaction at low temperatures.

\begin{figure}
\includegraphics[scale=1]{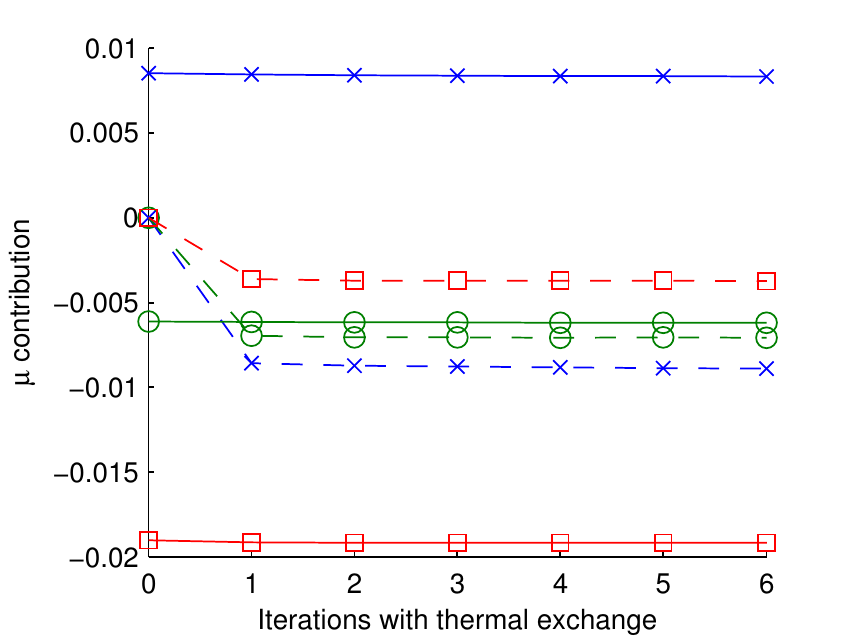}
\caption{\label{fig:thermalExMu}(Color online) Contribution to to the chemical potential (in units of $\hbar\omega_\rho$) from the direct (solid line) and exchange (dashed line) thermal dipolar interactions for a few iterations after thermal exchange is turned on. The trap aspect ratios used are $\lambda=\sqrt{8}$ (blue crosses), $\lambda=1$ (green circles) and $\lambda=1/\sqrt{8}$ (red squares). The other parameters used are $N=100$, $T=\hbar\omega_\rho/k_B$, $D=3.5$ and $\tilde{g}=0$.}
\end{figure}

We have examined the effect of the energy cutoff on the relative sizes of the direct and exchange thermal effects to determine how important different modes are to each effect. In Tab.~\ref{tab:muRatio} we show the ratio, $|\mu_E/\mu_{\tilde{D}}|$, between the thermal and direct dipolar exchange contributions to the chemical potential. As the cutoff energy is increased from four times to eight times the temperature, the relative importance of the exchange effect increases. This indicates that the higher energy, less occupied modes contribute more to the exchange effect than the direct effect. It is not clear that this trend will hold for higher energy modes which will become important at higher temperatures, however the HFBP calculations rapidly become infeasible as the energy cutoff is increased further.

In order to present results for a system where a significant fraction of the atoms are in the thermal cloud, we assume a very small condensate with only 100 atoms. This allows us to use a relatively low temperature, and hence energy cutoff, while keeping 1-10\% of atoms in the thermal cloud. This is enough for the thermal cloud to have a significant effect, but not so much as to require many iterations for self-consistency. We have performed calculations for three different trap aspect ratios: cigar-shaped, pancake-shaped and spherical. 

In Fig.~\ref{fig:thermalExMu} we show the contribution to the chemical potential from the direct and exchange thermal dipolar terms. We have plotted these for several iterations of the algorithm with the thermal exchange term included and we can see that there is reasonable convergence with only a few iterations. The contribution to the chemical potential from the direct thermal dipolar term is positive for a $\lambda=\sqrt{8}$ pancake trap as the repulsive part of the dipolar potential dominates in this trap. The contribution from the thermal exchange term is negative however, and of roughly the same magnitude. In a spherical trap the two terms both have a negative contribution of very similar magnitude. The size of both effects is reduced slightly compared to the pancake trap case. Finally, for the case of a cigar trap with $\lambda=1/\sqrt{8}$, the direct thermal dipolar contribution is strongly negative as the attractive part of the dipolar potential dominates. The thermal dipolar exchange term is again negative and smaller in magnitude than the previous cases. In the cigar trap, the size of the exchange term is much smaller than the direct effect.

\begin{figure}
\includegraphics[scale=1]{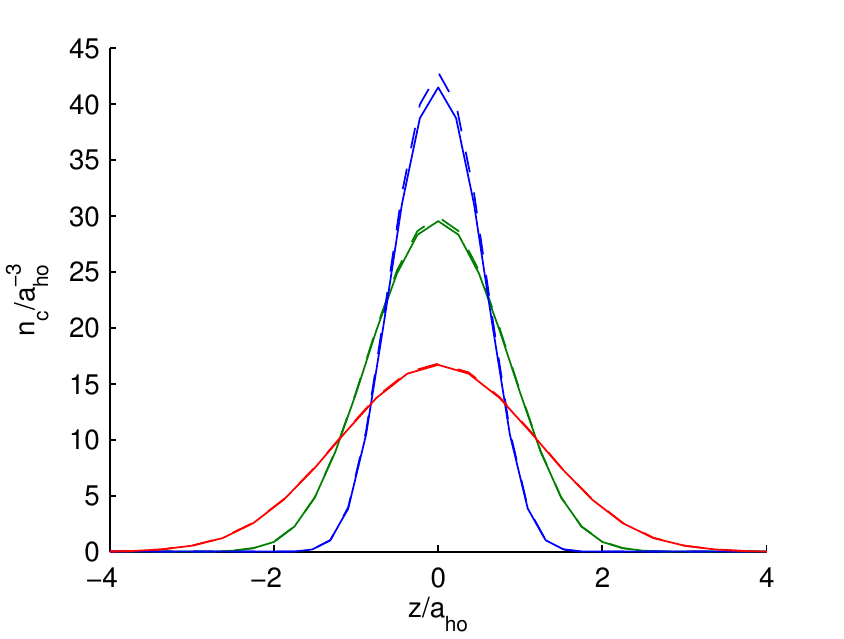}
\caption{\label{fig:thermExSlice}(Color online) Condensate density cross-section along the $z$-axis extrapolated to $\rho=0$. Parameters used are the same as in Fig.~\ref{fig:thermalExMu} with trap aspect ratios $\lambda=\sqrt{8}$ (blue), $\lambda=1$ (green) and $\lambda=1/\sqrt{8}$ (red). The density before thermal dipolar exchange is included is given by the solid lines, while the dashed lines show the density after convergence with thermal dipolar exchange.}
\end{figure}

In Fig.~\ref{fig:thermExSlice} we plot density cross-sections of the condensate for the same three cases. In each case, the central density is increased slightly with the inclusion of thermal dipolar exchange. The effect is strongest for the pancake shaped trap, with the difference becoming imperceptible on the scale of the figure for the cigar trap. This matches the relative sizes of the thermal exchange effect on the chemical potential for the three trap aspect ratios. The importance of the thermal dipolar exchange term appears to be quite dependent on the trap aspect ratio with the strongest effects in a pancake shaped trap. The thermal dipolar exchange term decreases the chemical potential and increases the central density of the condensate in each of the three cases shown. This shows that it is primarily an attractive interaction as opposed to the direct dipolar interaction which can be attractive or repulsive depending on the aspect ratio of the trap.

\begin{table}
\begin{tabular}{ccc @{\hskip 6pt}c}
\hline
\hline
$m$ & Parity & $E_j$ & $\Delta E_j$\\
\hline
0 & Odd & 1.00 & $4.9\times 10^{-3}$\\
0 & Even & 1.72 & $3.1\times 10^{-4}$\\
0 & Even & 2.15 & $1.2\times 10^{-2}$\\
1 & Even & 1.01 & $4.0\times 10^{-3}$\\
1 & Odd & 2.19 & $9.6\times 10^{-3}$\\
2 & Even & 2.39 & $8.7\times 10^{-3}$\\
\hline
\hline
\end{tabular}
\caption{\label{tab:Excits}Shifts in the excitation energies due to thermal dipolar exchange for a spherical trap with the same parameters as in Fig.~\ref{fig:thermalExMu}. The six lowest energy excitations are shown. The energies, $E_j$, are calculated without the thermal dipolar exchange term and $\Delta E_j$ is the shift in these energies when this term is included. Energies are in units of $\hbar\omega_\rho$.}
\end{table}

To show the effect that the thermal exchange interaction has on the excitation energies, we have calculated the shift in excitation energies when it is turned on. Tab.~\ref{tab:Excits} shows the excitation energies and their shifts  for a spherical trap. The shifts in energy are very small and this is to be expected as there is little shift in the excitation energies at small but finite temperatures even when thermal dipolar exchange is excluded \cite{Ronen2007b}. The two Kohn modes (with $E_j\approx1$) differ from the expected value slightly due to limitations of the HFBP method where the thermal cloud is static \cite{JPhysB}. The results for the cigar and pancake shaped traps are similar with the effects again being largest for the pancake trap.

The overriding conclusion however is that thermal exchange can be at least as important as the direct interaction. This indicates care should always be exercised when neglecting thermal exchange.

\section{Conclusions}
We have examined the effect of dipolar interactions on a harmonically trapped BEC. To begin with, we ignore the thermal dipolar exchange term. We have analysed the density profiles of the condensate and thermal cloud. In a spherically symmetric trap we can see how the anisotropic dipolar interactions break the spherical symmetry of these density profiles. We calculate the zero-temperature Bogoliubov excitations of the system for a wide range of trap aspect ratios. Due to the limitation on the number of modes we can feasibly include in the calculations, we are limited to very low temperatures. We find that the excitation energies shift very little for these temperatures. The dipolar interactions break the degeneracy between a number of modes.

We then include the thermal dipolar exchange interaction to determine its effect. This term cannot be expressed as an effective potential and it is much more computationally demanding to calculate than the direct dipolar term. We have calculated the effect of the thermal dipolar exchange term on the chemical potential for different trap aspect ratios. We find that the term lowers the chemical potential and is most significant for a pancake shaped trap. We also examine the effect on the density profile of the condensate. Again, the effect is largest for the pancake shaped trap and it tends to increase the central density. Finally, we determine what effect the thermal dipolar exchange term has on the excitation energies of the system. We find that the shift in the excitation energies due to thermal exchange is small.

Overall, we find that the effect of the thermal dipolar exchange can be as large as the thermal direct term at very low temperatures. These effects only become significant for very small condensates. Calculations for significantly higher temperatures become much more difficult as more modes need to be included. For the number of modes we have used (approximately fifty), calculations including exchange can take a few hundred hours.

\section{Acknowledgments}

Firstly, we would like to thank Danny Baillie for his considerable assistance, especially at the beginning of this project and during the writing of the manuscript. We would also like to thank Blair Blakie for his input in discussions. Financially this work was supported through contract NERF-UOOX0703 of the New Zealand Foundation for Research, Science and Technology and also by the National Research Foundation and the Ministry of Education of Singapore.

%

\end{document}